\documentclass[twocolumn,aps,prb,showpacs,floatfix,epsfig]{revtex4}

\usepackage{graphicx}%

\usepackage{dcolumn}

\usepackage{bm}%

\newcommand \be{\begin{equation}}

\newcommand \ee{\end{equation}}

\newcommand \ba{\begin{eqnarray}}

\newcommand \ea{\end{eqnarray}}

\begin{document}

\title
{\Large\bf On stochastic switching of 
bistable resonant-tunneling structures
via nucleation}

\author{Pavel Rodin$^1$\cite{EMAIL} and Eckehard Sch{\"o}ll$^2$}
\affiliation{$^1$Ioffe Physicotechnical Institute of Russian Academy of Sciences,
Politechnicheskaya 26, 194021, St.-Petersburg, Russia,\\
$^2$Institut f{\"u}r Theoretische Physik, Technische Universit{\"a}t Berlin,
Hardenbergstrasse 36, D-10623, Berlin, Germany
}

\setcounter{page}{1}

\date{\today}


\hyphenation{cha-rac-te-ris-tics}

\hyphenation{se-mi-con-duc-tor}

\hyphenation{fluc-tua-tion}

\hyphenation{fi-la-men-ta-tion}

\hyphenation{self--con-sis-tent}

\hyphenation{cor-res-pon-ding}

\hyphenation{con-duc-ti-vi-ti-tes}



\begin{abstract}

We estimate the critical size of the initial nucleus of the
low current state in a bistable resonant tunneling structure
which is needed for this nucleus to develop into a lateral switching
front. Using the results obtained for deterministic switching fronts,
we argue that for realistic structural parameters the critical
nucleus has macroscopic dimensions and therefore  is too large to be
created by stochastic electron noise.
\end{abstract}

\maketitle

In Refs.\ \onlinecite{MAT04,MAT03},  
the following switching mechanism has been discussed for
the  double-barrier resonant tunneling structure (DBRT) in presence of
electron shot noise. A bistable DBRT with Z--shaped
current--voltage characteristic is considered. The bistability is due to the
charge accumulation in the quantum well.\cite{GOL87a} The high current and
low current states correspond to the high and low electron concentration
$n$ in the quantum  well, respectively.
The high current state, which is stable in absence of
fluctuations, becomes metastable due to electron shot noise
when the voltage $V$ is chosen close
to the threshold voltage $V_{\rm th}$ at the upper boundary of
the bistability range (see Fig.\ 1 in Ref.\ \onlinecite{MAT03,MAT04}).
Hence the system eventually jumps to the low current state.
Whereas in small structures this occurs uniformly over the whole
area of the device,\cite{KOG98}
in large area structures the transition may occur via {\it nucleation}.
The nucleation is a two-stage process:
First the transition happens
in a small part of the device, forming an initial nucleus of the new
state. Then this initial nucleus expands,
leading to the transition of the whole structure to
the new state. ( This mechanism has been originally discussed with respect
to bistable microstructures in Ref.\ \onlinecite{KOG98}.)
However, in analogy with the well-known case of an equilibrium phase
transition, to enable expansion of the initial nucleus of the new state,
its lateral size $r$ should exceed
a certain critical size $r_{\rm cr}$. \cite{KOG98,MIK94,SCH01}
Consequently, a quantitative estimate of $r_{\rm cr}$  would be
useful in order to understand the relevance of the nucleation switching 
scenario in a DBRT.

The expansion of the nucleus represents a deterministic process
of switching front propagation.
Such nonlinear fronts in bistable DBRT have been studied in
Refs.\ \onlinecite{GLA97,MEL98,FEI98,MEI00,CHE00,SCH02,ROD03}.
Refs.\ \onlinecite{FEI98,MEI00,CHE00,SCH02,ROD03} are specifically
devoted to the sequential tunneling regime, considered
in Ref.\ \onlinecite{MAT03}.
The critical size $r_{\rm cr}$ can be estimated on the basis of these
results.  We shall focus on the lower bound for
$r_{\rm cr}$ which is given by the characteristic diffusion length
$\ell_{\rm D}$ of the
spatially distributed bistable system.\cite{KOG98,MIK94,SCH01}
Nuclei of smaller size $r < \ell_{\rm D}$
disappear  due the lateral spreading of the electron charge in the
quantum well and do not trigger a switching front.
The effect of curvature should also be taken into account
in case of two lateral dimensions as considered in 
Ref.\ \onlinecite{MAT03,MAT04},
when the initial nucleus is cylindrical. This generally
makes $r_{\rm cr}$ larger than $\ell_{\rm D}$.\cite{MIK94} However,
regardless of the lateral dimensionality $\ell_{\rm D}$ corresponds
to the lower bound for $r_{\rm cr}$.

Let us start with a simple analytical estimate for the front width.
For a bistable DBRT, $\ell_{\rm D} $ is determined by the lateral
spreading of the electron concentration in the well and the balance of
the emitter-well and well-collector currents in the vertical
(cathode-anode) direction which determines the regeneration of
the stored electron charge.
In the sequential tunneling regime
the lateral spreading of electrons in the quantum
well is dominated by an electron drift in the self-induced lateral
electrical field. \cite{CHE00,ROD03}
It is known from the theory of pattern formation
in active systems\cite{MIK94,SCH01}
that $\ell_{\rm D}$ is close to the width of the interface which
connects the coexisting on and off state in the stationary or moving
current density pattern such as a current density filament or
front. The order of magnitude of the velocity of the switching
front is given by (Eq.\ (19) in Ref.\ \onlinecite{ROD03})
\begin{equation}
v \sim \sqrt{\frac{\mu \Gamma_L E^F_e}{e\hbar}},
\end{equation}
where $\mu$ is the electron mobility in the well, $\Gamma_L/\hbar$
is the tunneling rate via the emitter-well barrier,
$E^F_e$ is the Fermi
level in the emitter, and $e$ is the electron charge.
The concentration of stored electrons in the front
can be roughly approximated by a piecewise
exponential profile (Eq.\ (A5),(A6) in Ref.\ \onlinecite{ROD03}).
The characteristic
``decay length'' $\ell_{W}$ of this profile is of the order of
\begin{equation}
\ell_W \sim \frac{\hbar}{\Gamma_L} \; 
v = \sqrt{\frac{\mu \hbar E^F_e}{e 
\Gamma_L}},
\end{equation}
as immediately follows from Eqs.\ (17),(19),(A5),(A6)
in Ref.\ \onlinecite{ROD03}.
The front width $W$, defined as 
the width over which the electron concentration changes by approximately 95\%
of the high-to-low ratio,
is related to $\ell_W$  as $W \approx 3 \ell_W$.
For typical values $\mu \sim 10^5 \, {\rm cm^2/V \, s}$,
$\Gamma_L \sim 1 \, {\rm meV}$,
$E^F_e \sim  10 \, {\rm meV} $ we obtain
$v \sim 10^7 \; {\rm cm/s}$ and
$W \sim 1 \; {\rm \mu m}$.

Eq.\ (2) gives the characteristic scale of the front width and
reveals its dependence on the main structural parameters.
A quantitive evaluation of $W$ follows from
numerical simulations \cite{ROD03} which show that $W$
can be larger than predicted by Eq.\ (2).
According to Ref. \onlinecite{ROD03} the front width is about
$10 \; {\rm \mu m}$ for a stationary front 
(the structural parameters: 
$\mu \sim 10^5 \, {\rm cm^2/V \, s}$;
$\Gamma_L \sim 0.5 \, {\rm meV}$,
$\Gamma_R \sim 0.1 \, {\rm meV}$), 
it increases with  voltage, i.e. for a moving front,
and becomes several times larger near the threshold voltage
$V_{\rm th}$, i.e. at the end of the range of bistability (see Fig.\ 4b in 
Ref.\ \onlinecite{ROD03}, where a 
stationary front ($v=0$) according to Fig.\ 4a corresponds to a 
voltage $|u| \approx 370 \, {\rm mV}$, and 
$V_{\rm th}=|u_{\rm th}| \approx 410 {\rm mV}$ according to 
Fig.\ 2 therein).
Note that the mobility $\mu$ in the well depends on the scattering
time and therefore is  related to the broadening of the
quasibound state in the quantum well.
Since the bistability range of the DBRT structure
shrinks and eventually disappears when the broadening of the
quasibound state increases, it is not possible
to substantially  decrease $\ell_{\rm D}$ by choosing
a low mobility $\mu$.

Another limitation imposed on $r_{\rm cr}$ is related to the 
vertical thickness of the tunneling structure $w$. 
The dependence of the energy of the quasibound state on the 
stored electron charge $\Delta E \ = e^2 n / C$
(Eq.\ 11 in Ref.\ \onlinecite{MAT03}, where $C$ is 
the effective capacitance of the well) is applicable only 
for variations of $n$ whose characteristic length $\lambda$
is much larger than $w$.\cite{GLA97,CHE00}
Variations of $n$ with $\lambda < w$ are screened by the electrons in
the highly conducting emitter and collector regions.
Since such variations  do not influence  the energy of the well and
therefore do not contribute to the electrostatic feedback mechanism
which leads to bistability, 
we obtain $r_{\rm cr} > w$. However, the condition 
$r_{\rm cr} > \ell_{\rm D}$
is much stronger because $\ell_{\rm D} \gg w$.

Our estimate $r_{\rm cr} > \ell_{\rm D} \sim W \sim 10 \; {\rm \mu m}$
suggests that for realistic DBRT parameters
the nucleus represents a macroscopic object whose lateral
dimension is comparable to the typical lateral size of a DBRT
structure.\cite{GOL87a}
Physically, this results from the efficient
re-distribution of electron charge in the quantum well plane.
Since the transition probability decreases exponentially with
the area of the nucleus,\cite{MAT03,KOG98}
the probability of spontaneous appearance of the critical nucleus due
to shot noise in the structure with extra-large area
$S \gtrsim \pi r_{\rm cr}^2$ is negligible.
We note that the probability of the stochastic generation of a critical
nucleus is equal to the probability that a DBRT with a lateral size of
$2 r_{\rm cr} \approx 20 \; {\rm \mu m}$ is uniformly switched
by electron shot noise.

In Ref.\ \onlinecite{MAT03} a characteristic scale
$r_0 = \sqrt{\eta}(\alpha \gamma)^{-1/4}$ was introduced, where
$r_0$ is a characteristic width of the critical profile $n(r)$
(this profile is shown in Fig.\ 3 in Ref.\ \onlinecite{MAT03}),
which corresponds to the
saddle point of the functional $F(n)$ (Eq.(14) in
Ref.\ \onlinecite{MAT03}). This width
is determined  by the coefficient of lateral diffusion in the well $D$
($\eta \sim D$) and the parameters of the effective potential $\alpha$ and
$\gamma$ which reflects the balance of the emitter-well and
well-emitter currents (Eq.\ (5) in Ref.\ \onlinecite{MAT03}).
The physical meaning of $r_0$ is similar to
the meaning  of $r_{\rm cr}$ in our consideration.
It is shown that
$r_0 \sim (V_{\rm th}-V)^{-1}$ and thus $\pi r_0^2 \gg S$
for $(V_{\rm th}-V) \rightarrow 0$.\cite{MAT03}
In this case only uniform transitions are possible.
The characteristic time of such uniform transitions is
given by Eq.\ (6) in Ref. \ \onlinecite{MAT03}.
Since $r_0$ decreases with increase of $(V_{\rm th} - V)$,
it is assumed that $\pi r_0^2 < S$ for sufficiently large
$(V_{\rm th}-V)$ and then the switching via nucleation becomes
possible.\cite{MAT03} Ref.\ \onlinecite{MAT03} does not
provide a lower bound for $r_0$ in this regime, implicitly
assuming that $r_0$ becomes sufficiently small.

In Ref.\ \onlinecite{MAT04}, which represents an extended 
and elaborated version of  Ref. \onlinecite{MAT03}, $r_0$
is replaced by another characteristic length $d$.
It is related to $r_{0}$ as  \cite{MAT04} 
$$
\left( \frac{d}{r_0} \right)^{4} = \frac{8 \zeta \eta \alpha}{\gamma},
$$
where $\zeta \approx  7.75$ (Eq.(57) in Ref.\ \onlinecite{MAT04}).
The length  $d$ is evaluated as \cite{MAT04}
\begin{equation}
d \sim \frac{1}{\sqrt n}
\left(\frac{\hbar \sigma}{e^2 T_R} \right)^{3/4}
\left(\frac{T_L}{T_R}  \right)^{1/2},
\label{d1}
\end{equation}
where $\sigma$ is the conductivity in the quantum well and 
$T_L$, $T_R$ are transition coefficients of the barriers.
For $\sigma = e  \,  \mu  \, n, T_L = T_R  \sim \Gamma_R / E^F_e$
and $n \sim \rho_0 \, E_e^F$ 
from (\ref{d1}) we obtain
\begin{equation}
d \sim \left(\frac{\mu \hbar E_e^F}{e \Gamma_R} \right)^{3/4}
(\rho_0 E_e^F)^{1/4} = 
\left(\frac{m \hbar}{\pi e^3 } \right)^{1/4} 
\left(\frac{\mu}{\Gamma_R} \right)^{3/4} E_e^F,
\label{d2}
\end{equation}
where $\rho_0 \equiv m/ \pi \hbar^2$ is the two-dimensional
density of states.
For $\mu \sim 10^5 \, {\rm cm^2/V \, s}$,
$\Gamma_L \sim 1 \, {\rm meV}$,
$E^F_e \sim  10 \, {\rm meV}$ 
we obtain $d \sim 1 \, {\rm \mu m}$, 
in agreement with our estimate based on Eq. (2).

Ref.\ \onlinecite{MAT04} suggests that the lower
bound for $d$ given by Eq.(3) is 
\begin{equation}
d \sim \frac{1}{\sqrt{n}} \sim \frac{1}{\sqrt{\rho_o E^F_e}}
\approx 20 \; {\rm nm}.
\end{equation}
This value corresponds  to 
$T_L = T_R \approx 1$ and the conductivity in the well  
$\sigma = e^2/ \hbar$. Both assumptions
are far beyond the limits of applicability of the model
under consideration.
Indeed, the analysis in Refs.\ \onlinecite{MAT04,MAT03}
is done for the incoherent tunneling regime
whereas the limit of transparent barriers $T_L = T_R \approx 1$ 
clearly corresponds to the coherent tunneling regime.
The conductivity $\sigma = e^2/\hbar \sim 2.4 \cdot 10^{-4} \; \Omega^{-1}$
corresponds to the low mobility limit when the mean free path
of an electron is of the order of the average distance between
carriers (effective mobility $\widetilde 
\mu = \sigma / (e n) \sim 5 \cdot 10^3 \; {\rm cm^2/Vs}$).
Strong electron scattering generally leads to wide 
broadening of the level in the quantum well, which is known to smooth out
the bistability of the resonant-tunneling structure.\cite{ROD03}
The broadening $\Gamma_{\rm well}$
of the quasibound state corresponding to $\sigma=e^2/\hbar$
is of the order of  $E^F_e$ which is far too large to observe 
bistability. Finally, $d = 20$~nm does not meet the 
condition $d > w$ for the typical width of the resonant-tunneling
structure $w \sim 100$~nm. We conclude that the evaluation of $d$
for $T_{L,R} \rightarrow 1$ and $\sigma  \rightarrow e^2/\hbar$ is 
physically meaningless.

Our consideration shows that for realistic structural parameters
the crossover from uniform stochastic 
switching to stochastic switching via nucleation 
might not happen because $r_{\rm cr}^2$ remains
comparable to $S$ regardless of the voltage $V$. In principle, the nucleation scenario
remains possible in extra-large structures ($S \gg \pi r_{\rm cr} ^2$)
when $(V_{\rm th} - V)$ is chosen sufficiently small so that
$S \gg \pi r_0^2 > \pi r_{\rm cr}^2$.
In practice, in this macroscopic limit the statistical properties of the
switching time are determined rather
by the fluctuations of the applied voltage $V$ and broadening of the
threshold voltage $V_{\rm th}$ due to imperfections of the DBRT
structure. The inaccuracy of the applied voltage becomes
particularly important for transient measurements like those
performed in Ref.\ \onlinecite{ROG01} when the applied
voltage is dynamically increased in a stepwise manner
to reach the metastable state at the edge of the bistability range.

In conclusion, our estimates based on the results 
of the studies of lateral switching fronts in a bistable DBRT
\cite{ROD03} suggest
that the  critical nucleus, which is needed to
trigger such a switching front,  has a macroscopically large lateral
dimension (e.g. $\ge 10 \; {\rm \mu m}$) for realistic structural 
parameters. 
Therefore it is doubtful that the nucleation
scenario  which implies triggering of the lateral front by
shot electron noise is possible in a DBRT.
The effect of electron shot noise on the lifetime
of the metastable state rather decreases with
increasing area of the DBRT structure and vanishes for structures with
macroscopic lateral dimensions,
in agreement with Ref.\ \onlinecite{KOG98}.
This does not exclude that the nucleation
mechanism might be relevant in other bistable systems.

We are grateful to A. Amann for useful discussion.
One of the authors (P.R.) thanks A. Alekseev
for the hospitality at the mathematical department
of the University of Geneva and  acknowledges support by the
Alexander von Humboldt Foundation
and Swiss National Science Foundation.

\end{document}